# Guidelines for the Creation of an Annotated Corpus

Bahdja Boudoua[1,2], Nadia Guiffant[1,2], Mathieu Roche[1,3],
Maguelonne Teisseire[1,2], Annelise Tran[1,2,4]

1 TETIS, Univ. Montpellier, AgroParisTech, CIRAD, CNRS, INRAE, Montpellier, France
2 INRAE, UMR TETIS, Montpellier, France
3 CIRAD, UMR TETIS, F-34398 Montpellier, France
4 UMR ASTRE, Univ. Montpellier, CIRAD, INRAE, Montpellier, France

## Abstract

This document, based on feedback from UMR TETIS members and the scientific literature, provides a generic methodology for creating annotation guidelines and annotated textual datasets (corpora). It covers methodological aspects, as well as storage, sharing, and valorization of the data. It includes definitions and examples to clearly illustrate each step of the process, thus providing a comprehensive framework to support the creation and use of corpora in various research contexts.

## Introduction

Best practices for creating an annotated corpus cover the entire data lifecycle: collection, processing, analysis, storage, sharing, and reuse. This guide includes recommendations based on the feedback from members of the UMR TETIS and the literature. Its goal is to synthesize the process of creating an annotation guideline, producing textual datasets (corpora), and managing their storage.

## 1 Corpus

### 1.1 Definition of the annotated corpus

According to the definition provided by the the National Center for Textual and Lexical Resources (Centre National de Ressources Textuelles et Lexicales):

- In linguistics, a corpus is a collection of texts compiled based on a framework of exhaustive documentation, a thematic or exemplary standard for linguistic study.
- In data science, a corpus refers to a set of data that can be utilized in an analysis or automatic information retrieval experiment.

Annotating a corpus involves adding one or more layers of linguistic interpretation to the data within the corpus. Annotations are carried out during annotation campaigns that bring together several human annotators, with different levels of expertise, who rely on an annotation guideline.

The creation and evaluation of annotated corpora, as well as automatic annotation methods, are receiving increasing interest in linguistics and Natural Language Processing (NLP), particularly due to the growing development of Machine Learning applications.

In the context of this guide and the examples provided, the annotation is done at the document level. Other approaches, such as sentence-level or word-level annotation, are also possible. An example of these different types of annotation in the field of epidemiology is illustrated in Figure 1 below: the selected news article is successively annotated at the document, sentence, and word levels.

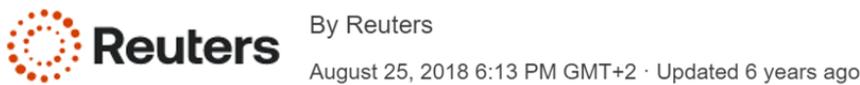

**African Swine Fever hits Romania's biggest pigs farm**

"We've been focusing on mainland and the virus might have emerged from the waters," he said. Romania has reported hundreds of outbreaks of the disease among **pigs** kept in backyards and smallholdings as well as several large private farms located especially in the south of the country.
About 100,000 pigs have been culled so far.
African swine fever affects **pigs** and **wild boar** and has spread in Eastern Europe in recent years. It does not affect humans.

Figure 1: Snippet from a Reuters article. The full article is available at the following address: https://www.reuters.com/article/us-romania-swineflu-pigs-idUSKCN1LA0LR/.

| Annotation type | Label / Example |
|---|---|
| Document level | Relevant (Epidemiological outbreaks' description) |
| Sentence level (Highlighted sentence) | Assessment / Consequence |
| Word level (Pigs, Boars) | Affected host |

Table 1: Possible Types of Annotations Applied to the Example Illustrated in Figure 1.

## 1.2 Data collection

For the creation of a corpus, existing data (texts, snippets) are collected, often from the Internet. It is important to justify the choice of the collected data (in this case, textual documents) with regard to the research objectives.

### 1.3 Development of the Annotation guidelines

The development of the annotation guidelines generally follows an iterative process (Figure 2) (Sabou et al. 2014). This process involves several annotation rounds conducted on a sample by experts, with the goal of producing a precise guide that enables annotators, even non-experts, to annotate the articles without running into ambiguity issues.

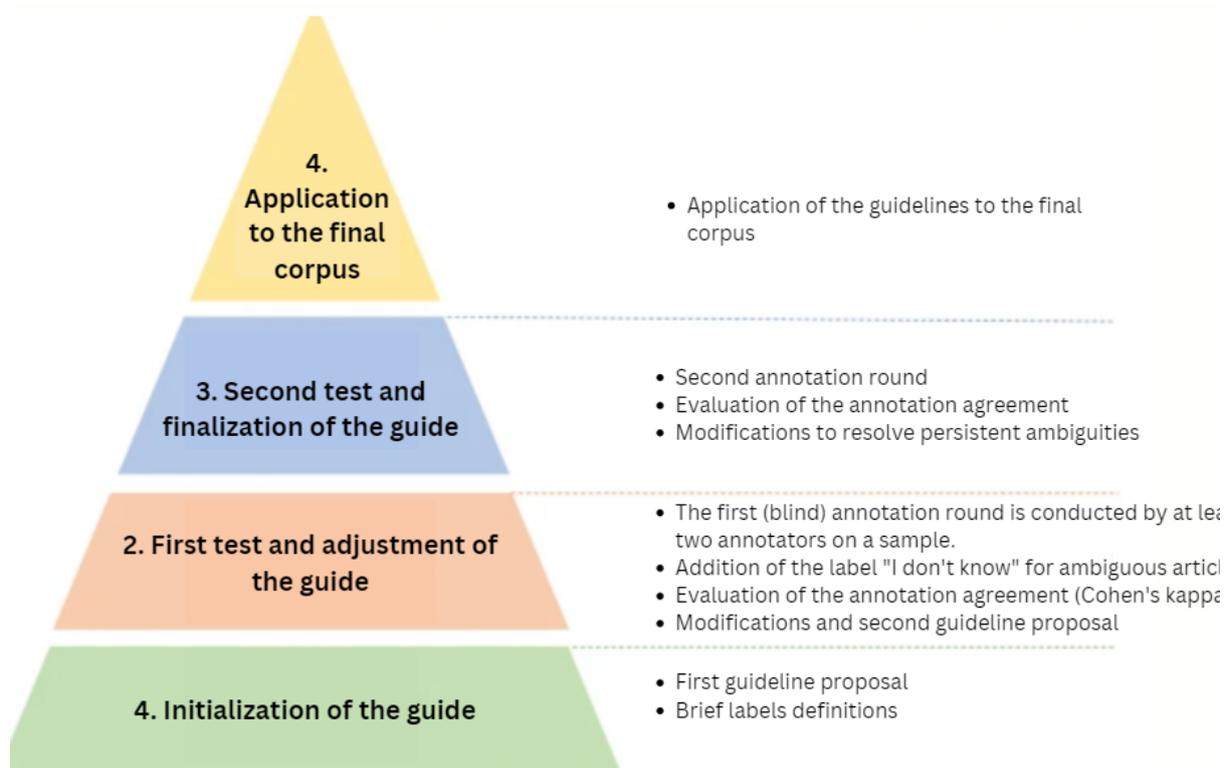

Figure 2: Annotation guidelines elaboration process (adapted from Valentin et al. 2022)

#### 1.3.1 Development of a preliminary version of the guidelines

The first step involves establishing an initial version of the annotation guide, with concise and clear definitions of the annotation categories.

**Note:** It is important to include the label 'I don't know' in order to highlight documents or ambiguous phrases that are difficult to interpret. This option helps prevent annotators from feeling forced to classify articles (or sentences/segments) into inappropriate categories.

#### 1.3.2 Blind Annotation Phase

Once this preliminary version is established, a representative sample of the dataset to be annotated (for example, 40 to 50 articles) is selected. This sample is then annotated independently and blindly by at least two experts. This step allows for testing the consistency of the definitions.

After this step, the inter-annotator agreement is calculated. Several methods can be used, such as Cohen's Kappa coefficient (McHugh 2012), which is appropriate when two annotators are involved, and Fleiss's Kappa (Zheng et al. 2017), which is used when more than two annotators are involved.

### 1.3.3 Analysis and discussion of disagreements

The disagreements observed in the annotations are then discussed among the experts. This discussion allows for the identification of ambiguities or imprecisions in the definitions of the annotation categories. There are no strict rules; adjustments depend on the dataset. For example, it may be necessary to add new categories or to merge existing classes.

### 1.3.4 Revision and second round of annotation

After the revision of the guideline, a new annotation phase is conducted by the same experts. This second round allows for checking that the adjustments made to the annotation guide address the issues. Based on experience, two rounds of annotation are usually sufficient, but it is essential to rely on the agreement among annotators to assess the effectiveness of this process and, if necessary, to organize additional rounds.

### 1.3.5 Finalization of the annotation Guidelines

When the annotation agreement reaches a satisfactory level, the final version of the guideline is established. A Kappa value of 0.6 is considered acceptable, while 0.9 is regarded as perfect. Generally, a Kappa value of 0.8 is recommended as the minimum acceptable threshold for inter-annotator agreement (McHugh, 2012).
The final guidelines will then be used to annotate the entire dataset.

In summary, this work results in the production of an annotation guide that, once applied to the corpus, generates an annotated dataset. These data must then be stored and organized to enable their effective utilization and re-utilization. The next section addresses the storage and sharing of data.

## 2 Storage and sharing of the data

### 2.1 Data Repository

Once the annotated corpus has been created, it is essential to make it accessible and reusable. Assigning a persistent identifier (PID) or a unique identifier (DOI) is recommended to enable reliable location and citation of the data. These identifiers also facilitate tracking citations and reuse.

A reliable and long-term repository automatically assigns an identifier to each dataset. Data deposit in a repository is free of charge. At a minimum, a repository should contain the annotated data along with the associated annotation guidelines. It is also possible to store different versions of the data, such as raw data and annotated data, to ensure better traceability of their development.

Since 2021, the Research Data Gouv repository has been recommended by ministerial authorities.

There are also [thematic repositories](), some specifically dedicated to corpora. For the TETIS unit, it is particularly recommended to use the Dataverse (CIRAD)[1] and DataGouv (INRAE)[2] platforms.

You can deposit your data and complete their metadata without necessarily publishing them immediately. For instance, this could be in the case of an embargo if the data is awaiting publication by a journal or if you believe the data still needs modifications before being shared.

When publishing data online, choosing a Creative Commons license specifies the terms of use and gives you control over how others may reuse your data. There are several types of licenses, ranging from the most permissive to the most restrictive. More information is available here.

*Note*: The collected data is protected by licenses or General Terms of Use (GTU) issued by its producers. Various types exist, specifying the rights granted for reuse (e.g., free or paid access, rights to reproduce, copy, modify, extract, share, redistribute, publish, etc.). To legally use the data, it is essential to verify that their licenses or GTUs allow the intended processing.

**Tip:** There are approaches you can adopt to work around these restrictions while respecting licenses and GTUs

| Do's | Don't s |
| --- | --- |
| To share tweets without mentioning user IDs | To share tweets in full |
| To share texts snippets | Share full texts without permission |
| To share URLs to articles | Share online media articles without permission |

### 2.2 Documenting the data

Metadata are the data or information used to describe your research data (e.g., title, date, author, project context, keywords, etc.). They help in understanding the data and knowing their origin; they are essential for facilitating the use of the data since they often represent the only means of communication between the stages of data production and secondary analysis. Therefore, they must be clear and provide all the useful information for the analysis and reuse of the data.

The figure below presents the general metadata associated with a dataset produced as part of the MOOD - News AMR dataset - Hackathon 2022 (Arınık et al. 2022).

---

[1] https://dataverse.cirad.fr/
[2] https://entrepot.recherche.data.gouv.fr/dataverse/inrae

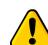

Figure 3: General metadata associated with the dataset produced in the context of the MOOD - News AMR dataset - Hackathon 2022.

⚠️ Consider the documentation that would be necessary to enable the reuse of your data. This may include information about the methodology used to compile the corpus, the procedures and methods of analysis employed, the definition of variables (variable dictionary), units of measurement, and more.

Proper storage not only ensures the security and sustainability of the data but also facilitates its accessibility and future reuse. To ensure optimal management of the produced data, it is recommended to follow the FAIR principles (Findable, Accessible, Interoperable, Reusable), which define best practices for data management.

## 3  Sharing and promoting research data

Publishing in the form of a **data paper** is a preferred and recommended option for sharing data formally and informing the scientific community of their availability. A data paper is a **scientific article designed to describe and document a dataset.** Published as open access in a traditional scientific journal or a data journal (a journal dedicated exclusively to data papers), it is characterized by a specific structure that emphasizes **the collection methodology and technical analyses that validate the quality of the data.** Several resources, such as the 'Où Publier' (Where to publish)[3] database from CIRAD, list specialized journals for data papers to facilitate their dissemination within the scientific community.

**Examples of data papers and annotation guidelines produced at UMR TETIS**

1. Valentin, S., Arsevska, E., Vilain, A. *et al.* Elaboration of a new framework for fine-grained epidemiological annotation. *Sci Data* **9**, 655 (2022). https://doi.org/10.1038/s41597-022-01743-2

2. Arınık, N., Van Bortel, W., Boudoua, B., Busani, L., Decoupes, R., Interdonato, R. Teisseire, M. (2023). An annotated dataset for event-based surveillance of antimicrobial resistance. *Data in Brief*, *46*

3. Koptelov, M., Holveck, M., Cremilleux, B. *et al.* A manually annotated corpus in French for the study of urbanization and the natural risk prevention. *Sci Data* **10**, 818 (2023). https://doi.org/10.1038/s41597-023-02705-y

4. Holveck, Margaux; Koptelov, Maksim; Roche, Mathieu; Teisseire, Maguelonne, 2023, "Lisez_Moi.pdf", *Segments textuels - Textual Segments - Hérelles Project*, https://doi.org/10.57745/KT6JAB

5. Boudoua, El Bahdja; Richard, Manon; Roche, Mathieu; Teisseire, Maguelonne; Tran, Annelise, 2023, "AI_Annotation Guideline.pdf", Annotated datasets from PADI-web for event-based surveillance of Avian Influenza, African Swine Fever, and West-Nile Virus Disease, https://doi.org/10.57745/X3IZT3

---

[3] https://coop-ist.cirad.fr/fr/gerer-des-donnees/publier-un-data-paper/4-choisir-la-revue

**Appendix: Legal and Ethical Considerations**

As a reminder, the French Charter of Ethics for Research Professions[4] promotes, among other principles:

- freedom of expression
- independence of research
- reproducibility of research (proper referencing and access to raw and processed data)
- data storage limited to essential needs, in order to reduce environmental impact

In addition, dissemination is prohibited for certain types of data, including protected or sensitive scientific data, business intelligence data (industrial and commercial secrets), and data subject to statistical confidentiality.

Access to this data may, however, be granted to authorized individuals, on the condition that the data is stored in a secure location with controlled access and that any restrictions on its use are clearly specified.

Further information on the legal and ethical aspects of research data management is available at: https://doranum.fr/aspects-juridiques-ethiques/

Moreover, specific guidance on the processing of health data is provided in the CNIL's recommendations, available at the following address: https://www.cnil.fr/fr/quelles-formalites-pour-les-traitements-de-donnees-de-sante

**Acknowledgements**

This work was conducted in the context of the MOOD project (MOnitoring Outbreak events for Disease surveillance in a data science context; https://mood-h2020.eu/), funded by the European Union's Horizon 2020 programme under Grant Agreement No. 874850.

---

[4] https://comite-ethique.cnrs.fr/wp-content/uploads/2020/01/2015_Charte_nationale_d%C3%A9ontologie_190613.pdf